\documentclass[twocolumn, aps]{revtex4}
\usepackage{natbib}
\usepackage{graphicx}

\begin{document}

\title{Global divergence of microbial genome sequences
mediated by propagating fronts}
\author{Kalin Vetsigian and Nigel Goldenfeld}

\affiliation{Department of Physics and Institute for Genomic Biology,
University of Illinois at Urbana-Champaign, 1110 West Green Street,
Urbana, IL 61801-3080.}

\date{\today}

\begin{abstract}

We model the competition between homologous recombination and
point mutation in microbial genomes, and present evidence for two
distinct phases, one uniform, the other genetically diverse.
Depending on the specifics of homologous recombination, we find
that global sequence divergence can be mediated by fronts
propagating along the genome, whose characteristic signature on
genome structure is elucidated, and apparently observed in
closely-related {\it Bacillus} strains.  Front propagation
provides an emergent, generic mechanism for microbial \lq\lq
speciation", and suggests a classification of microorganisms on
the basis of their propensity to support propagating fronts.

\end{abstract}

\maketitle

The transfer of genetic material between microbial cells plays a
crucial role in their evolution, and poses fundamental questions
to microbiology. Is there a tree of life for microbes \cite{Woese,
Kurland, Doolittle}? Are there bacterial species
\cite{Lawrence_speciation_without_species, Cohan}? What are the
mechanisms driving their diversification \cite{Ochman, Doolittle,
Berg_Kurland, Redefining_bacterial_populations_review}? These
questions arise because genetic transfer couples the evolution of
different genomes in a way that not only complicates their
dynamics but obscures their very identity over time: the evolution
is communal. While in sexual organisms the communality of genome
evolution is restricted to species, the major elements of
microbial evolution---genetic transfer followed by illegitimate or
homologous recombination, point mutations, genome
rearrangements---do not {\it a priori} imply sharp genetic
isolation boundaries. If there are none, notions such as species
and speciation, despite being widely used heuristically, are
misleading. Also, it is not clear whether there are classes of
microbes with qualitatively different modes of communal evolution
and what are the cellular properties that distinguish between
them.

Gene transfer results when foreign DNA is taken up from the
environment (transformation), delivered by a virus (transduction)
or acquired via a direct cell to cell exchange (conjugation), and
then permanently incorporated in the recipient genome by
homologous or illegitimate recombination. Homologous
recombination, mediated by dedicated cellular machinery, plays a
vital error correction role in genome replication
\cite{recombination_and_replication} but also allows a foreign DNA
fragment to replace a sufficiently similar portion of the
recipient genome.  The probability of successful replacement in
homologous recombination is proportional to the exponential of the
number of sequence mismatches \cite{Vulic}, the mechanism being
organism-specific \cite{Majewski_Cohan, Majewski_Cohan1,
Majewski_Cohan3}. Illegitimate recombination can be mediated by
bacteriophage integrases, selfish genetic elements, or occur by
chance DNA breakage and repair, and allows the acquisition of
entirely novel traits from evolutionary distant organisms.
Illegitimate genetic transfer, also known as horizontal gene
transfer (HGT), can be inferred from the genome data through its
atypical sequence composition \cite{Ochman} and the phylogenetic
incongruences it causes \cite{Doolittle1}. While the extent of HGT
is under heated debate \cite{Kurland}, it is clear that it is much
less frequent than homologous recombination. Relative rates of
homologous recombination and point mutations in natural
populations have been estimated by sequence diversity studies
using multi-locus sequence typing data in recently-formed
bacterial strains \cite{Guttman_Dykhuizen, Feil_Spratt}. The
probability that a gene changes as a result of homologous
recombination can be many times higher than that for point
mutations. Another manifestation of the pervasiveness of
homologous recombination is that the evolution of strains within
many named species cannot be represented by a phylogenetic tree
\cite{How_clonal_are_bacteria, Feil_Spratt_PNAS, MILK97}.
While the importance of genetic transfer, and homologous
recombination in particular, is firmly established
\cite{Feil_Nature}, there are only a few sharp predictions about
the resulting modes of microbial evolution. Relevant to our work
is the observation of Lawrence
\cite{Lawrence_speciation_without_species} that HGT islands
locally inhibit recombination. He concludes that global genetic
isolation can be achieved through the gradual accumulation of
hundreds of HGTs.

The purpose of this paper is to explore the emergent properties of
the collective evolution of closely related bacterial genomes. We
model the interplay of homologous recombination and point
mutations in bacterial populations, and show that elementary
genome changes such as HGT, genome rearrangements, insertions or
deletions can trigger diversification fronts that in evolutionary
short time propagate along the bacterial genomes and eventually
lead to global sequence divergence of sub-populations. The
diversification fronts can occur even in the absence of natural
selection and demonstrate that fast neutral evolution can have
non-trivial long-term evolutionary consequences. The robustness of
this mechanism is sensitive to some of the details of homologous
recombination, and suggests a way to classify the spectrum of
evolutionary modes in bacteria based on specific details of their
homologous recombination mechanisms. We establish a methodology
for analyzing closely related genomes and give evidence for a
large-scale step-like variation of homologous recombination rates
in the {\it Bacillus cereus\/} group, which might be a signature
of a diversification front. Finally, we discuss the biological
implications of the propagation of diversification fronts, as a
mechanism for speciation, a force favoring the formation of sharp
genetic isolation boundaries, and a dynamical barrier for HGT and
genome rearrangements.

The details of homologous recombination are by now reasonably well
understood \cite{Vulic, Majewski_Cohan}. There are at least two common
obstacles to successful integration of a DNA fragment. First, the end
of the fragment must find a short region ($\approx 20$ bp) of sequence
identity with the target genome in order to initiate the process.
Second, the cell's mismatch repair system can abort the recombination
process if it encounters mismatches between the fragment and the
portion of the genome being replaced. Both of these obstacles lead to
an exponential decrease of recombination with sequence divergence.
There are also potentially important variations in the mechanism. While
in {\it E. coli} sequence identity at only one end is required, in {\it
Bacillus} very high sequence similarity at both ends is needed
\cite{Majewski_Cohan, Majewski_Cohan1} and mismatch repair seems less
important. In {\it Streptococcus} the effect of mismatch repair is
intermediate in strength \cite{Majewski_Cohan3} but the overall
dependance of sexual isolation on sequence divergence is very close to
that in {\it Bacillus}. In addition, the underlying bases for
distinguishing between donor and recipient DNA can differ. Do these
differences in the details translate into qualitatively different
evolutionary behavior? If so, then the details of the homologous
recombination mechanism could be an important criterion for classifying
bacteria. The computational studies described here clarify which
details are the relevant determinants of the long-term evolutionary
dynamics.

\medskip
\noindent {\it Models:-} Based on the above considerations, we construct
sets of model rules that describe the interplay between homologous
recombination and point mutations.
\begin{enumerate}
\item There are $N$ circular strings of length $L$ written in an
alphabet of $n$ symbols.
\item Each position in each genome is subject to point mutations
with rate $m$. A point mutation changes a symbol to any other
symbol with equal probability.
\item  Each genome receives fragments at an average rate $r$. Each
fragment is of size $F$, is derived from an arbitrary position
from an arbitrary donor genome and attempts to recombine at the
same genome position in the recipient.
\item To be considered for incorporation the fragment must find an
identical segment of length $M$ at an arbitrary chosen end (Model I) or at
both ends (Model II).
\item The probability of incorporation is $\exp(-\alpha d)$, where
$\alpha$ is a coefficient expressing the strength of the mismatch
repair system and $d$ is the pointwise sequence difference, i.e.
$d$ counts the number of mismatches between the fragment and the
genome sequence it is about to replace.  We will also consider
Model III, where rule 4 is absent.
\end{enumerate}
The genome strings can be thought of as representatives of different strains
possessing at least partial ecological distinctiveness, so that random
genetic drift is much stronger within strains than between strains.
With this interpretation we do not include random genetic drift but it
can be straightforwardly added.

\medskip
\noindent
{\it Propagation of diversification fronts:-} In these models,
mutation and recombination play opposing roles: point mutations
generate sequence diversity in the population, whereas
recombination tends to make sequences more similar. At high
recombination rates an initially uniform population will remain
close to uniform; at high mutation rates all sequences will
diverge from each other. An important property of homologous
recombination is that the probability that a recombination event
is successful decreases with sequence divergence and becomes
negligible even for small levels of divergence \cite{Vulic}.

\begin{figure}

\includegraphics[width=\columnwidth]{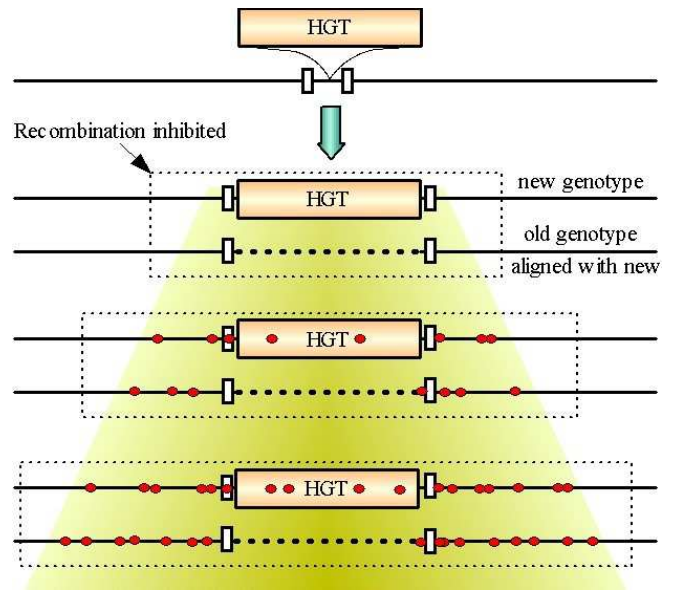}
\caption{Schematic illustrating the process by which a
diversification front propagates along a genome in a selection
neutral situation. In the vicinity of the HGT island,
recombination is suppressed relative to point mutations, allowing
point differences to build up in the region flanking the HGT
island. The newly accumulated sequence differences lead to the
extension of the region where recombination is inhibited and, in
turn, an accumulation of point differences further away from the
HGT island. The process repeats itself.}
\label{front_prop_cartoon}
\end{figure}
These considerations suggest that the uniform phase is {\it
metastable}: even when recombination is strong enough to maintain
a state of near uniformity it will not succeed in bringing
together sufficiently diverged sequences. The diverged phase on
the other hand is stable. If there is a boundary between a stable
and a metastable phase the generic expectation is that the stable
phase will grow at the expense of the metastable one, as shown in
Fig. \ref{front_prop_cartoon}. This will happen because homologous
recombination is inhibited not only in the diverged phase but also
in a finite region flanking it within the uniform phase. Mutations
will accumulate in the flanking region, and as a result the
diverged phase will grow.   We will refer to the boundary between
the uniform and diverged phases as a {\it diversification front}.
Therefore, the system has the potential to sustain the propagation
of diversification fronts. Such diversification fronts can be
nucleated by processes that create regions of sequence difference
between genomes in the population, such as HGT, genome
rearrangements, deletions or insertions and have important
biological consequences for the evolution and diversification of
microbes, as will be discussed later.

\medskip
\noindent {\it Simulations:-} To clarify this intuition we
performed a series of simulations of a population of interacting
genomes, starting from two different initial conditions: 1) all
sequences are the same, and 2) all sequences are the same except
for a strip, long compared with the typical size of recombining
fragments, in which the sequences are random. We used three
different models for the rules governing the dynamical behavior of
homologous recombination: Model I, requiring sequence identity at
one end of the recombining fragment; Model II, requiring sequence
identity at both ends; and Model III, with no requirement of
sequence identity. The central questions addressed are: Under what
circumstances is there a well defined front propagation region; is
it readily observable or is fine tuning of the parameters
required? Do the three models differ qualitatively?  To address
these questions in a quantitative manner, we define an order
parameter
\begin{equation}
\psi(x) = \frac{n}{n-1}\frac{1}{N(N-1)}\sum_{i,j}
{(1-\delta_{A_{xi}, A_{xj}}}) \label{eqn:order_parameter}
\end{equation}
where $A_{xi}$ denotes the letter at position $x$ of genome $i$.
The order parameter $\psi$ measures the average difference in the
population between the sequences at genome position $x$ normalized
so that $\psi=1$ when the genomes are uncorrelated.  This
corresponds to the {\it diverged phase\/} of the system.  In the
opposite limit, $\psi=0$, the genomes in the system are highly
correlated, giving rise to the {\it uniform phase\/} of the
system.

For each model, we studied the time evolution of the order
parameter for different values of $m/r$ and $\alpha$. Typical
values used for the other parameters are $F=500$, $M=10$,
$L=10000$, $N=20$, $n=2$. For each separate run we measured $\psi$
as a function of position within the genome and time.  By varying
$\alpha$, we control the strength of the mismatch repair
mechanism, and hence the success rate of recombination. The most
important trend probed by our simulations is the behavior of the
order parameter as a function of the ratio $\mu\equiv m/r$, the
relative strength of point mutations versus recombination.

\begin{figure}
\includegraphics[width=\columnwidth]{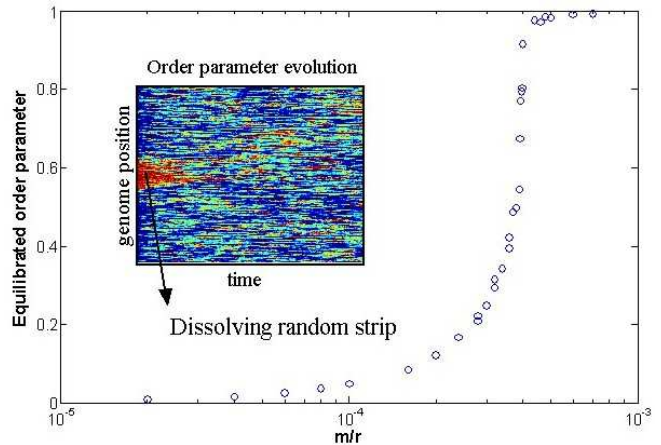}
\caption{The equilibrium value of the order parameter changes
gradually with $m/r$ in Model I with $\alpha=0$, $F=500$, $M=10$,
$L=10000$, $N=20$ and $n=2$. The inset figure depicts a typical
time evolution of the genome population. The vertical axis
represents position along the genome, the colorscale indicating
the value of the order parameter (blue denoting uniform phase, red
denoting diverged phase), while the horizontal axis is simulation
time.  A random strip dissolves without triggering a
diversification front.} \label{gradual_transition}
\end{figure}

\medskip
\noindent {\it Results for Models I and III:-} For sufficiently
low values of $\alpha$, the equilibrium value of the order
parameter varies gradually with $\mu=m/r$, as shown in Fig.
\ref{gradual_transition}. The uniform and random strip initial
conditions always relax to the same final state. The random strip
simply dissolves and no front propagation is observed.  This
situation arises when recombination is allowed almost regardless
of the degree of sequence divergence.

\begin{figure}
\includegraphics[width=\columnwidth]{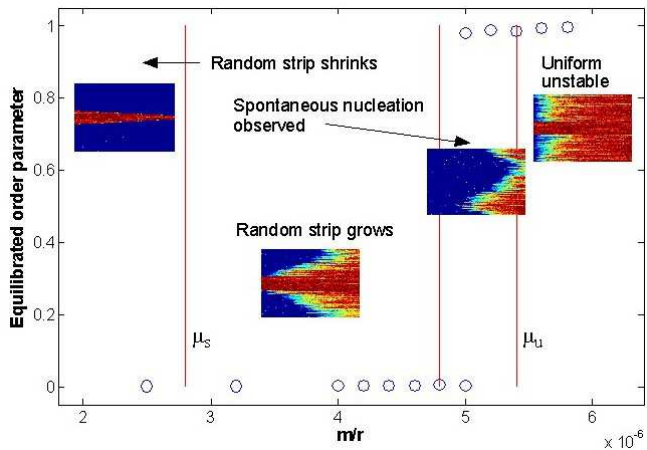}
\caption{Starting from a uniform state, the order parameter
equilibrates to values close to 0 or 1 in Model I with
$\alpha=0.4$, $F=500$, $M=10$, $L=10000$, $N=20$ and $n=2$,
indicating the existence of distinct uniform and diverged phases.
The inset figures depict the genome population for the indicated
value of $m/r$, as a function of time.  The vertical axis
represents position along the genome, the colorscale indicating
the value of the order parameter (blue denoting uniform phase, red
denoting diverged phase), while the horizontal axis is simulation
time. For $\mu_s < \mu < \mu_u$ the random strip triggers a
diversification front. For $\mu$ close to $\mu_u$ spontaneous
nucleation is possible. }\label{sharp_transition}
\end{figure}

Above a threshold value of $\alpha$, the uniform and diverged
phases become distinct: for small values of $\mu$, the order
parameter is 0, and the system is genetically uniform. However,
for large values of $\mu$, the order parameter is close to unity,
indicating that the system is genetically diverged. This
transition appears to be sharp, as shown in Fig.
\ref{sharp_transition}.  There is further interesting dynamical
behavior as a function of $\mu$. For $\mu > \mu_u$ the uniform
phase becomes unstable and the sequences diverge everywhere
simultaneously.  For $\mu < \mu_s$, the uniform phase is stable,
and a finite region of diverged phase shrinks as a function of
time, i.e. the uniform phase invades the diverged one. For $\mu_s
< \mu < \mu_u$, diversification proceeds through nucleation and
growth of the diverged phase; in this parameter range, front
propagation occurs.

\begin{figure}
\includegraphics[width=\columnwidth]{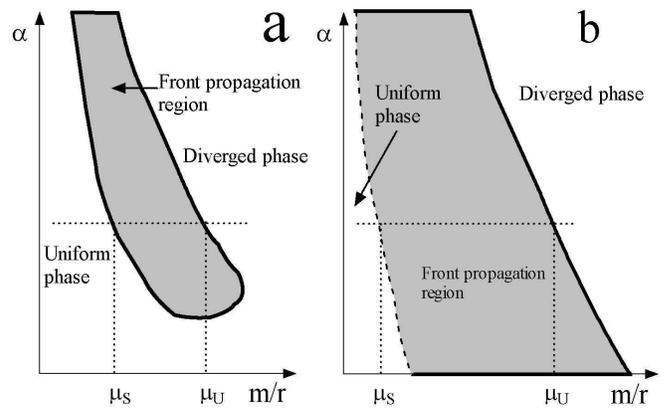}
\caption{{\bf a.} The phase diagram of Models I and III. Distinct
phases exist only above a threshold value of $\alpha$ and the
width of the front propagation region, $\mu_u/\mu_s$, is less than
2. {\bf b.} The phase diagram of Model II. Distinct phases exist
for all values of $\alpha$ and the front propagation region is
very wide: $\mu_u/\mu_s>100$.}\label{phase_diag_1}
\end{figure}

From this behavior, we deduce the qualitative phase diagram presented
in Fig.~\ref{phase_diag_1}a. Model III, with no sequence identity
requirement, shows qualitatively similar results (data not shown).

\medskip
\noindent {\it Results for Model II:-}  For Model II, with
sequence identity requirement at both ends, we observe front
propagation even for $\alpha=0$. Moreover, the width $w\equiv
\mu_u/\mu_s$ of the interval $\mu_s < \mu < \mu_u$, where front
propagation occurs, is very wide.  While for Models I and III we
always observed $w \le 2$, for Model II we could not even observe
the point $\mu_u$, and $w > 100$. This results in the phase
diagram qualitatively represented on Fig.~\ref{phase_diag_1}b.
The front speed can be as high as several times the fragment size per
average point mutation time near the transition, and is a rapidly
decreasing function of the recombination rate.

To summarize, there is a qualitative difference between the
situation with no sequence identity requirement (Model III) or
sequence identity requirement at only one end (Model I) and Model
II with sequence identity requirement at both ends. The difference
is manifested in the phase diagram and the width of the front
propagation region.

\medskip
\noindent {\it Microbe classification:-} These theoretical
predictions imply that we can classify microbial genomes according
to the details of the recombination dynamics: class I, consisting
of models I and III, and class II, consisting of model II.  The
distinguishing feature of the classes is whether or not the
recombination dynamics requires sequence identity at both ends of
the incorporated segment. For Class II, as long as the uniformity
of a population is maintained by homologous recombination, it will
support propagating diversification fronts. For Class I,
diversification fronts are possible only within a narrow interval
of the ratio of mutation to recombination rates and are therefore
unlikely.

The existence of class I and class II indicates that the details
of homologous recombination are important beyond the fact that the
probability of recombination exponentially decreases with sequence
divergence. Therefore it is necessary to elucidate further the
differences between homologous recombination mechanisms in
different bacteria and work out their consequences for front
propagation. For example, if mismatch repair is nick-directed and
not methyl-directed \cite{Majewski_Cohan3} then more mismatches
will be detected near the ends of the recombining fragments. This,
in turn, will make front propagation more robust, because a
greater fraction of the average homogenizing capability of
recombination will be inhibited by a phase boundary. Also, if
non-homologous DNA loops formed during the recombination process
are not corrected efficiently, then small deletions, insertions,
slippage and inversions would not trigger diversification fronts.
Since micro rearrangements are presumably frequent, the efficiency
of loop repair will be an important factor in determining the rate
of nucleation of fronts. Finally, it is important to know whether
or not and how the length of the incorporated fragments is
dynamically dependant on the differences between the donor and
recipient.

In order to seek evidence for the front propagation mechanism, we
now compare available completely sequenced genomes of
closely-related microbes. The most direct evidence for front
propagation from genome data alone would be an extended step-like
pattern in the sequence divergence of closely-related well-aligned
genomes, with the diverged region centered around a region of HGT,
deletion or genome rearrangement. The front profile reflects the
different times after genetic isolation of different parts of the
chromosome. Under conventional uniform molecular clock
assumptions, it will be approximately linear, with a slope
determined by the distance the front travels during the time it
takes the sequences to fully diverge once recombination is
inhibited. Slowly changing components of the sequence divergence,
such as non-synonymous substitutions, leads to more extended
profiles.

\medskip
\noindent {\it Analysis of genome data:-} We consider the sequenced
genomes in the genus {\it Bacillus}. It is in {\it Bacillus\/} that
Majewski and Cohan \cite{Majewski_Cohan1} discovered the requirement
for sequence identity at both ends, and our simulations indicate that
front propagation is more likely to occur in such systems.

We obtained the complete genome sequences from the NCBI database,
together with the positions and orientations of the known or
predicted protein coding regions, tRNAs and rRNAs.  We globally
aligned all pairs using the {\tt nucmer} script of the MUMMER
package \cite{MUMMER} ({\tt nucmer -b 50 -g 300 -c 65 -mum}), obtaining
a list of well aligned regions for each pair. Three {\it Bacillus
cereus\/} strains - ATCC 10987, ATCC 14579 and ZK
\cite{Cereus_Nature, Cereus2}, three {\it Bacillus anthracis}
strains - Ames, Ames Ancestor and Sterne, and {\it Bacillus
thuringiensis serovar konkukian str. 97-27} genomes were close,
highly co-linear and analyzed further. The three anthracis strains
were practically identical and only Ames was used in the analysis.

\begin{figure}
\includegraphics[width=\columnwidth]{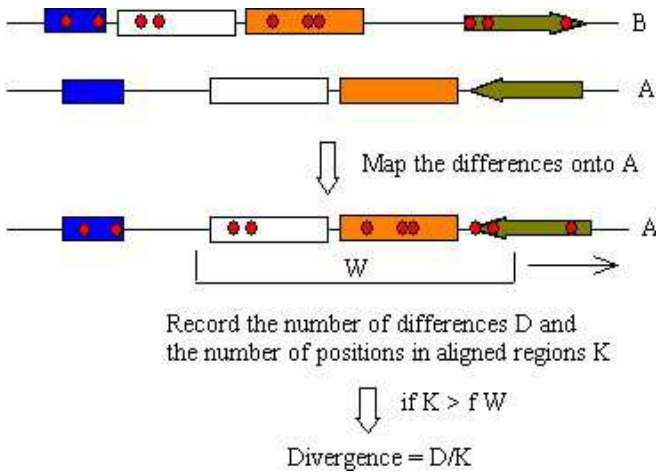}
\caption{To construct the divergence profiles we first identify
the well aligned regions (represented by color bars and arrows)
using MUMMER, then map the differences (represented by red
circles) onto the reference genome and slide a window of width $W$
along the genome.}\label{coarse_graining}
\end{figure}

For each pair, we mapped the well-aligned regions on one of the
genomes, and constructed a series of coarse-grained profiles by sliding
a window of width $W$ along the genome while excluding non-aligned
regions (resulting from insertions and deletions) from the averaging,
as depicted graphically in Fig.~\ref{coarse_graining}.
The profiles have gaps where the window covers less than a
threshold fraction $f$ of $f W$ unambiguously aligned nucleotides.
We used $W$ in the range of 40k to 120k and $f$ between 0.5 and
0.8. We looked at the coarse-grained profiles for the DNA point
differences, as well as intergene, intragene, 3rd codon, 1st and
2nd codon, synonymous and non-synonymous (as defined in
\cite{Ks_Ka}) differences.

{\it Cereus ATCC 10987} exhibits a distinct step-like pattern of
sequence difference when compared to {\it cereus ZK}, {\it
antracis ames} and {\it thuringiensis serovar konkukian str.
97-27}. The pattern is also present in each of the other
difference components - synonymous, non-synonymous, gene,
intergene. What is the explanation for this pattern? Does it
involve homologous recombination or not? Is it a result of a front
propagation during the separation of {\it cereus ATCC 10987} with
the common ancestor of {\it cereus ZK}, {\it antracis ames} and
{\it thuringiensis serovar konkukian str. 97-27}?

To answer these questions, we first examined the variation of the
nucleotide composition along the genome. Based on the GC and AT
skews the replication terminus is located at around 2.6Mb -- away
from the position of the difference profile step. The GC content
varies smoothly along the genome and does not exhibit a step
pattern. It has a minimum near the replication terminus.

The step pattern is partially correlated with the density of
protein coding regions in the above genomes, the sequence
differences being larger where the density is lower. However,
since all difference components  exhibit the pattern, it cannot be
simply an artifact due to different proportions of gene and
intergene regions with different mutation rates. Moreover, within
the well aligned regions, the intergene regions are, on average,
only about 15\% more divergent than protein coding regions and the
gene density varies only in the 75-90\% percent range. Therefore,
the small differences in the proportions of sites with different
mutation rates would have to have been somehow amplified if
varying coding density were the underlying cause of the pattern.
The non-aligned regions have a higher intergene fraction than
aligned ones suggesting a possible mechanism by which the density
of protein coding regions can indirectly affect sequence
divergence by a preferential accumulation of inter-strain
alignment gaps in intergene regions and a corresponding reduction
of recombination rates.

Could it be that not just the proportion of site types, but the point
mutation rates themselves vary gradually along the genome, leading to
the above pattern? To answer this question, we turn to the distribution
of lengths of maximal exact matches (DLMEM) between pairs of aligned
sequences. If differences had accumulated by a Poisson mutational
process, then we would expect an exponential distribution. Recombination, on
the other hand, will lead to a broader distribution and, for example, a deviation
from the Poisson statistics value (unity) for the ratio of the standard
deviation and the mean \cite{Sawyer}.

Whether these deviations are statistically significant can be
determined by comparing with the distribution of this ratio for
the case without recombination.

\begin{figure}
\includegraphics[width=\columnwidth]{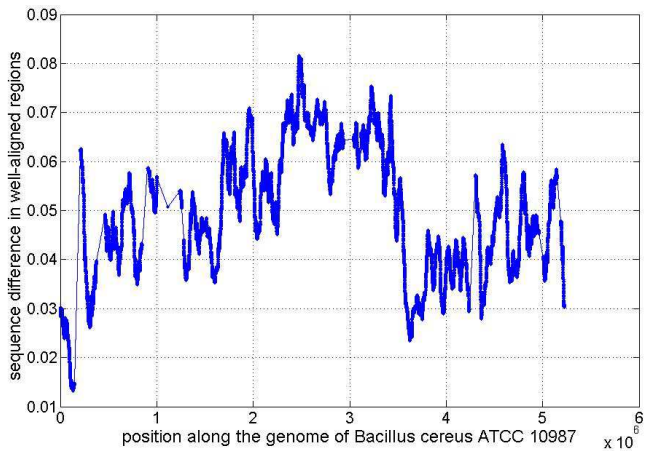}
\caption{The step-like profile of the sequence difference between
{\it Bacillus cereus ATCC 10987} and {\it Bacillus cereus ZK}
obtained by sliding a 60k window with $f=2/3$ along the genome.}
\label{The_pattern}
\end{figure}
\begin{figure}
\includegraphics[width=\columnwidth]{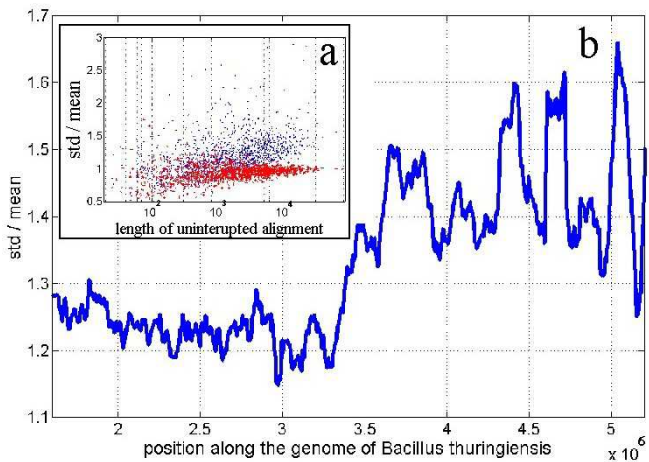}
\caption{DLMEM statistics resulting from the comparison of {\it
Bacillus thuringiensis} and {\it Bacillus cereus ATCC 10987}. {\bf
a.} The std/mean for the distribution of lengths of maximal exact
matches within a well-aligned region is positively correlated with
the length of the region. The actual data (blue dots) is
contrasted with a null hypothesis with matched sequence difference
for each region (red *) {\bf b.} The std/mean DLMEM profile
obtained using a 120k window with $f=0.5$ along {\it Bacillus
thuringiensis} exhibits a step-like
pattern.}\label{Std_mean_Regions}
\end{figure}

We gathered DLMEM statistics for different well-aligned regions.
The ratio of the standard deviation and mean is significantly
above 1, as shown in Fig.~(\ref{Std_mean_Regions}a). Moreover, there is a
positive correlation between this ratio and the length of the
uninterrupted well-aligned regions, a trend which agrees with the notion
that non-aligned parts inhibit recombination within the adjacent
aligned regions.

We then looked for evidence of different rates of homologous
recombination along the chromosome by studying the changes in the
DLMEM statistics in a sliding window. There is again a step-like
pattern for the ratio of the standard deviation and the mean, as shown
in Fig.~(\ref{Std_mean_Regions}b).

Deviation of the ratio of the standard deviation and the mean of a
DLMEM is a sign of clustering of the differences along the
chromosome. Are there reasons for clustering which do not involve
homologous recombination? If different genes have very different
evolution rates, then this can lead to apparent clustering. For
example, different gene expression levels can lead to different
synonymous mutation rates and an apparent clustering of
differences within the weakly expressed genes. To control for
this, we compare the DLMEM for neutral mutations with a null model
with matched neutral divergence of each protein coding region
separately. The pattern is present in the real data but almost
completely disappears in the control. The residue is due to
correlations of the divergences of adjacent proteins which are
expected in the presence of homologous recombination. Since,
presumably, there is no reason apart for recombination for
clustering of synonymous substitutions within each gene
separately, this test not only rules out genes with different
evolutionary rates as an explanation but also gives confidence
that the standard deviation over mean deviations from unity are
predominantly due to homologous recombination.

Further evidence supporting the homologous recombination
interpretation of the ratio of the standard deviation and the mean
of DLMEM comes from contrasting the above observations with the
results of the comparison between the completely sequenced {\it
Buchnera aphidicola} strains {\it APS, BP and SG}. Because, these
are intracellular parasites lacking the RecA gene we expect no
homologous recombination. Indeed, we find that there is no
statistically significant deviation from unity of the standard
deviation over mean and a highly uniform difference profile.

In summary, the above data indicate that there are large-scale
step-like variations of the rates of homologous recombination along the
analyzed microbial genomes, apparently consistent with the hypothesis
that diversification proceeded by front propagation.

\medskip
\noindent {\it Discussion:-} In this section, we discuss the
consequences of the front propagation mechanism for the fate of
bacteria that have acquired useful skills through HGT or have undergone
a large-scale genome rearrangement. We argue that the front propagation
mechanism facilitates global genetic isolation between strains, and, as
such, is a mechanism for what may be loosely termed \lq\lq speciation". On the
other hand, the front propagation mechanism reduces the chances that
chromosomal changes, such as incorporation of HGTs or rearrangements,
will be evolutionary successful, thus creating a dynamical barrier to
the accumulation of such mutations in evolutionary time.

A bacterium can acquire a new skill by means of HGT. This can lead to
the extinction of those bacteria which do not possess the beneficial
(under appropriate selection pressure) HGT fragment. Alternatively,
HGT can allow the invasion or foundation of a new biochemical niche,
while being disadvantageous in the former one, or lead to specialization
within the old niche.  (Indeed, ecological distinctiveness without
spatial isolation is not unusual for microbes. Even in the simplest of
environments - mono culture lab experiments - coexisting strains emerge
spontaneously \cite{spont_niche_formation}. However, the creation of
coexisting genotypes by HGT cannot properly be termed speciation, because the
genotypes are not genetically isolated with respect to homologous
recombination, except for a small region surrounding the HGT.)

The front propagation mechanism makes local isolation unstable, because
the HGT event nucleates a diversification front leading eventually to a
global isolation of the carriers of the HGT event from the rest of the
population. Therefore, ecological distinctiveness accompanied by local
isolation is enough to generate speciation, even when homologous
recombination is not reduced by the ecological distinctiveness.  Note
that this outcome is different from the one proposed by Lawrence
\cite{Lawrence_speciation_without_species}, who suggested that global
isolation is only achieved through the accumulation of hundreds of
HGTs.  Our work has demonstrated that even a single HGT or genome
rearrangement can lead to global sequence divergence.

It is difficult to apply the biological species concept to groups
of strains that are isolated at some loci and not at others
\cite{Lawrence_adolescence}. Because of diversification front
propagation, a community of bacteria in which pairs of bacteria
are genetically isolated at some loci, but not others, is unstable
and tends to partition itself into groups which are globally
isolated from each other with respect to homologous recombination.
This is because genetically isolated regions will suppress
recombination and trigger fronts into neighboring non-isolated
regions. This instability will be even stronger if the different
genomes are not colinear or do not have the same set of genes.
Therefore, well defined genetic isolation boundaries emerge
spontaneously through the front propagation mechanism even if
there is no functional barrier to gene transfer.

What happens when a HGT or a rearrangement brings some advantage,
but without enabling the recipient to adopt an entirely distinct
ecological role? Achieving complete ecological distinctiveness might
be a gradual process. In this case the new genotype will be
successful initially but not necessarily in the long run because
it will be competing with other beneficial mutations at other loci
that emerge throughout the population. Beneficial mutations
trigger selective sweeps that can be either global, purging the diversity
throughout some ecological niche or, because of homologous
recombination, local, purging the diversity only around the locus
of the beneficial mutation. In a population in which relative
sequence uniformity is maintained by homologous recombination,
local selective sweeps will be the norm. However, front
propagation, nucleated in the carriers of a HGT or a rearrangement
will propagate by accumulation of neutral mutations, and
potentially lead to global genetic isolation  of the carriers long
before they have a chance to achieve a full ecological
distinctiveness.

New strains are easily formed by readily absorbing foreign genetic
material, rearranging the genomes, etc. But they are typically
short-lived entities, because following front propagation they are
excluded from the communal evolution. Front propagation implies that
the evolutionary rate of HGT accumulation is less than the rate
suggested by looking at strains. This can be, in principle, tested
against the data. This mechanism can also explain why gene order is
highly conserved in some bacterial groups: there exists a dynamical
barrier to the survival of rearranged genomes.

These considerations also have implications for the applicability of
molecular phylogenetics, and the ongoing debate about the nature of the
impact of HGT on the tree of life.  Front propagation limits the impact
of HGT, reinforcing in a complementary way Woese's concept of a
complexity barrier to HGT \cite{Woese}.  Our argument is complementary,
because it does not rely on the nature of the interactions between the
genes: there is a barrier to HGT arising from the population dynamics
alone.

Our work leaves open a number of interesting issues related to the
effect of highly conserved regions on front propagation.  A large
immutable region can present an impassable obstacle to front
propagation. Candidates for such obstacles are rRNA operons, tRNA
genes and overlapping genes. Such regions lack the flexibility
arising from the degeneracy of the genetic code.  HGTs islands
inserted near front obstacles will lead to the diversification of
a smaller fraction of the recipient genome, and have a greater
chance to avoid extinction. Is there a correlation between
evolutionary persistent HGTs and RNA gene positions? If a genome
region is already diversified there is no penalty for the
incorporation of another useful HGT island. Is there clustering of
HGT islands? How is front propagation modified for clonal bacteria
\cite{MILK97}? Finally, is front propagation beneficial? If front
propagation obstacles are allowed to evolve or at least reposition
themselves, what configuration of obstacles would result?

On the basis of computer simulations, we have suggested that the
interplay between homologous recombination and point mutations can
lead to propagating fronts, in whose wake a population of microbes
becomes genetically diverse in evolutionary short time.  Thus,
even in the absence of selection pressure and ecological barriers
to genetic exchange, gene-exchange boundaries can emerge as a
statistical consequence of the detailed dynamics of recombination.
We have presented a preliminary analysis of available genome data
for the {\it Bacillus cereus} group, which is consistent with the
presence of front propagation. These findings prompt speculations
about the implications for the evolution and the classification of
microbes.

Our model can be extended in a number of directions, including explicit
accounting for the role of space, the existence of a non-trivial
network of gene exchange connectivity and the effects of sharing of
beneficial mutations.

A promising approach to looking for diversification fronts is
metagenomics data. Such data can give us a consensus genome for an
ensemble of closely related organisms, inhabiting the same
environment, and an estimate for the sequence diversity along the
consensus genome \cite{Phil}. This diversity can be directly
related to the order parameter $\psi(x)$. A step like variation in
$\psi(x)$ might be an indication of a diversification front.

We thank Phil Hugenholtz for bringing the work of Lawrence to our
attention after the main results of our study had been obtained and to
Yoshi Oono for useful discussions.  We also thank two anonymous
referees for helpful suggestions that improved this work.  This work
was partially supported by the National Science Foundation through
grant number NSF-EAR-02-21743.

\end{document}